\documentclass[a4paper]{IEEEtran}
\usepackage[boxruled]{algorithm2e}
\usepackage{enumerate}
\usepackage{cite}
\usepackage{graphicx}
\usepackage{psfrag}
\usepackage{subfigure}
\usepackage{url}
\usepackage{amsmath}
\usepackage{amsthm}
\usepackage{array}
\usepackage{amssymb}
\usepackage{amsfonts}
\usepackage{float}

\newtheorem*{conjecture*}{Conjecture}
\newtheorem{lemma}{Lemma}

\newtheorem{corollary}{Corollary}
\newtheorem{remark}{Remark}
\newtheorem{definition}{Definition}
\newtheorem{theorem}{Theorem}
\newtheorem{example}{Example}

\title{Rate $\frac{1}{3}$ Index Coding: Forbidden and Feasible Configurations}
\begin{document}
\author{
\IEEEauthorblockN{V. Lalitha and Prasad Krishnan\\}
\IEEEauthorblockA{SPCRC, International Institute of Information Technology, Hyderabad\\
Email: \{lalitha.v, prasad.krishnan\}@iiit.ac.in\\}
\vspace{-0.7cm}
}
\date{\today}
%\author{
%\authorblockN{K. Prasad}
%\authorblockA{Dept. of ECE, Indian Institute of Science \\
%Bangalore 560012, India\\
%Email:prasadk5@ece.iisc.ernet.in
%}
%\and
%\authorblockN{B. Sundar Rajan}
%\authorblockA{Dept. of ECE, Indian Institute of Science, \\Bangalore 560012, India\\
%Email: bsrajan@ece.iisc.ernet.in
%}
%}
\maketitle

\thispagestyle{empty}	
\pagestyle{empty}

%%%%%%%%
\begin{abstract}
Linear index coding can be formulated as an interference alignment problem, in which precoding vectors of the minimum possible length are to be assigned to the messages in such a way that the precoding vector of a demand (at some receiver) is independent of the space of the interference (non side-information) precoding vectors. An index code has rate $\frac{1}{l}$ if the assigned vectors are of length $l$.
%Necessary and sufficient conditions for symmetric rate $\frac{1}{3}$ index codes were obtained in recent work using the idea of interference configuration (non side-information) of special kind, known as type-2 alignment sets. 
In this paper, we introduce the notion of strictly rate $\frac{1}{L}$ message subsets which must necessarily be allocated precoding vectors from a strictly $L$-dimensional space ($L=1,2,3$) in any rate $\frac{1}{3}$ code. We develop a general necessary condition for rate $\frac{1}{3}$ feasibility using intersections of strictly rate $\frac{1}{L}$ message subsets. We apply the necessary condition to show that the presence of certain interference configurations makes the index coding problem rate $\frac{1}{3}$ infeasible. We also obtain a class of index coding problems, containing certain interference configurations, which are rate $\frac{1}{3}$ feasible based on the idea of \textit{contractions} of an index coding problem. Our necessary conditions for rate $\frac{1}{3}$ feasibility and the class of rate $\frac{1}{3}$ feasible problems obtained subsume all such known results for rate $\frac{1}{3}$ index coding.
\end{abstract}
%%%%%%%%%
\section{Introduction}
Index Coding, introduced in \cite{BiK}, considers the problem of efficiently broadcasting a set of messages available at a source, to a collection of receivers each of which already has a subset of the messages (called \textit{side-information}) and demands certain other subset of messages. Based on the configuration of the messages available as side-information and the demand sets, the general index coding problem is classified into various types including unicast \cite{BBJK} (with disjoint demand sets), uniprior \cite{OnHL} (with disjoint side-information sets), and the most general \textit{groupcast} index coding problems (arbitrary side-information and demand sets).  The goal of index coding is to find optimal index codes, where optimality means that the number of channel uses is minimum. To determine the maximum rate (inversely related to minimum number of channel uses) of an index coding problem is NP-hard. The landmark paper \cite{BBJK} famously connected the scalar linear index coding problem to finding a quantity called \textit{minrank} associated with the \textit{side-information graph} related to the given single unicast index coding problem. The minrank problem is NP-hard too, but several approaches have been taken to address this problem, most popularly via graph theoretical ideas to bound the maximum rate (see, for example, \cite{BiK,BBJK,SDL,BKL1,BKL2,TOJ}). The techniques used in these works to derive bounds on maximum rate naturally lead to constructions of (scalar and vector) linear index codes.

We focus on scalar linear codes in this paper. In \cite{Jaf,MCJ2} the index coding problem has been reformulated in the interference alignment framework, where constructing a scalar linear code was shown to be equivalent to assigning precoding vectors to the messages. The precoding vectors are assigned such that the space spanned by vectors assigned to interfering messages is linearly independent of the vector assigned to the demanded message. An index code is said to be a rate $\frac{1}{L}$ code (for some positive integer $L$) if the messages are scalars (from some field $\mathbb{F}$) and precoding vectors are $L$-length. In \cite{Jaf}, a necessary and sufficient condition for index codes of rate $\frac{1}{2}$ was presented based on the structure of the interference (messages not available as side-information), which is modeled using a \textit{conflict graph}. Consequently, a polynomial time algorithm to identify the feasibility of rate $\frac{1}{2}$ (or equivalently, minrank two) for a given index coding problem was given in \cite{Jaf}. 

 Unlike the rate $\frac{1}{2}$ feasibility problem which has a polynomial time solution,  the rate $\frac{1}{3}$ feasibility problem was shown to be NP-hard \cite{Pee}, when the finite field is fixed. A simple necessary condition for rate $\frac{1}{3}$ feasibility  was given in \cite{BBJK}. A class of rate $\frac{1}{3}$ feasible index coding problems was shown in \cite{Jaf}. In \cite{PrL}, a stricter necessary condition was derived than what was previously known in \cite{BBJK}. In addition, a larger class of index coding problems which are rate $\frac{1}{3}$ feasible was presented, which include those given in  \cite{Jaf}. Both the necessary and sufficient conditions obtained in \cite{PrL} are based on the following two ideas (i) \textit{conflict hypergraphs}, which preserve all the required information in the index coding problem (as opposed to conflict graph) (ii) \textit{type-2 alignment sets}, which are special subsets of messages which must necessarily be assigned vectors from a two dimensional space. 
 
In this work, we make further progress on the rate $\frac{1}{3}$ index coding problem. Our contributions are as follows.
\begin{itemize}
\item \textit{Strictly rate $\frac{1}{L}$ feasible subsets:} We introduce the notion of strictly rate $\frac{1}{L}$ feasible subsets of messages of an index coding problem ${\mathbb I}$. A subset of messages is said to be strictly rate $\frac{1}{L}$ feasible if for any rate $\frac{1}{3}$ solution, there are always $L$ linearly independent vectors amongst the precoding vectors assigned to the subset (Definition \ref{def:strict_rate} in Section \ref{sec:nec_condition} makes this precise). 
\item \textit{Necessary condition based on strictly rate $\frac{1}{L}$ feasible subsets:} We show that a certain method of `stitching' strictly rate $\frac{1}{2}$ feasible subsets together generates larger message subsets which should be strictly rate $\frac{1}{2}$ feasible. (Theorem \ref{thm:converse}, Section \ref{subsec:nec_condition}). Using this result, we recapture the main result of \cite{PrL}, while significantly generalizing it to include new configurations which must be strictly rate $\frac{1}{2}$. This, in turn, results in a general necessary condition for rate $\frac{1}{3}$ feasibility.
\item \textit{Contraction of an index coding problem:} We develop the notion of a \textit{contraction} of an index coding problem ${\mathbb I}$. A contraction of an index coding problem ${\mathbb I}$ is another index coding problem ${\mathbb I}'$ such that any solution for ${\mathbb I}'$ gives a solution for ${\mathbb I}$ (Section \ref{sec:suff_condition} gives a formal definition).
\item \textit{Sufficient condition based on contractions:} We give sufficient conditions on rate $\frac{1}{3}$ feasibility of ${\mathbb I}$ based on the structure of type-2 alignments sets of a maximal contraction of ${\mathbb I}$ (Theorem \ref{thm:suff_condition}, Section \ref{sec:suff_condition}). These conditions result in a larger class of problems which are rate $\frac{1}{3}$ feasible. 
\end{itemize}
%%%
%Random coding approaches to index coding were studied in \cite{ABKSW}. 
%Bounds on the rate of groupcast index coding were presented in \cite{TDN}. %
%The general class of \textit{groupcast} index coding problems consists of $n$ messages generated at a source, where each message is demanded by at least one receiver. 
%%%
\textit{Notations:} Throughout the paper, we use the following notations. Let $[1:m]$ denote $\{1,2,...,m\}$. For a set of vectors $A$, $sp(A)$ denotes their span. For a vector space $V$, $dim(V)$ denotes its dimension. An arbitrary finite field is denoted by $\mathbb F$. A vector from the $m$-dimensional vector space ${\mathbb F}^m$ is said to be picked \textit{at random} if it is selected according to the uniform distribution on ${\mathbb F}^m$.
%\vspace{-0.1cm}
\section{Review of Index Coding} \label{sec:review}

%The typical index coding problem involves commuication between one source and multiple receiver via a broadcast channel, in the presence of some prior information already available at the receivers. 
Formally, the general index coding problem, called a \textit{groupcast} index coding problem, consists of a broadcast channel which can carry symbols from $\mathbb F$, along with (i)  A set of $T$ receivers
(ii) A source which has messages ${\cal W}=\{W_i, i\in[1:n]\}$, each of which takes values from $\mathbb F$, (iii) For each receiver $j$, a set $D(j)\subseteq {\cal W}$ denoting the set of messages demanded by the receiver $j$, 
(iv) For each receiver $j$, a set $S(j)\subseteq {\cal W}\backslash D(j)$ denoting the set of side-information messages available at the $j^{th}$ receiver.
%%
%This general class of index coding problems is known as \textit{groupcast} index coding problems. %The more widely discussed class of problems is \textit{unicast}, where each message is demanded by a unique receiver.
%%
%\vspace{-0.2cm}
%%
%\begin{definition}[Scalar Linear Index code of symmetric rate $\frac{1}{L}$]
\textit{A scalar linear index code of symmetric rate} $\frac{1}{L}$ (for some integer $L\geq 1$) for a given index coding problem consists of an encoding function 
$
\mathbb{E}:\underbrace{\mathbb F\times\mathbb F\times...\times\mathbb F}_{n~\text{times}}\rightarrow {\mathbb F}^L,
$
mapping the messages ($W_i\in{\mathbb F}$) to some $L$-length codeword which is broadcast through the channel, as well as decoding functions at the receivers which use the codeword and the available side-information symbols to decode the demands. The encoding function of a scalar linear index code can be expressed as ${\mathbb E}(W_1,W_2,...,W_n)=\sum_{i=1}^nV_iW_i$,
where $V_i$ is a $L$-length vector over $\mathbb F$ called the precoding vector assigned to $W_i$. Finding a scalar linear index code of length $L$ (i.e., with a feasible rate $1/L$) is equivalent to finding an assignment of these $V_i$s to the $n$ messages such that the receivers can all decode their demanded messages.

\begin{definition}[Interfering sets and messages, conflicts]
For some receiver $j$ and for some message $W_k \in D(j)$, let $Interf_k(j)\triangleq {\cal W}\backslash({W_k\cup S(j)})$ denote the set of messages (except $W_k$) not available at the receiver $j$. The sets $Interf_k(j), \forall k$ are called the \textit{interfering sets at receiver} $j$. If receiver $j$ does not demand message $W_k$, then we define $Interf_k(j)\triangleq\phi$. If a message set ${\cal W}_i$ is not available at a receiver $j$ demanding at least one message $W_k\notin {\cal W}_i$, then ${\cal W}_i$ is said to \textit{interfere at receiver} $j$, and ${\cal W}_i$ and $W_k$ are said to be \textit{in conflict}.
\end{definition}

For a set of vertices $A\subseteq {\cal W}$, let $V_{\mathbb E}(A)$ denote the vector space spanned by the vectors assigned to the messages in $A$, under the specified encoding function $\mathbb E$. If $A=\phi$, we define $V_{\mathbb E}(A)$ as the zero vector. A message subset $A$ is said to be $L$-dimensional (under the code $\mathbb E$) if $dim(V_{\mathbb E}(A))=L.$
\begin{definition}[Resolved conflicts]
For a given assignment of vectors to the messages (or equivalently, for a given encoding function $\mathbb E$), we say that all the conflicts \textit{are resolved} if $V_k\notin V_{\mathbb E}(Interf_k(j)), \forall W_k\in D(j), \forall j$.
\end{definition}
It can easy to show that successful decoding at the receivers is possible if and only if all the conflicts are resolved \cite{PrL}. We alternatively refer to a collection of messages by only their indices (for example, $\{W_1,W_2\}$ is referred to as $\{1,2\}$).
%%%%
\section{Prior results from \cite{PrL} on Rate $\frac{1}{3}$ Index Codes} \label{sec:prior}
%In \cite{PrL}, the index coding problem was represented by a \textit{conflict hypergraph} (which itself was a development of the idea of a conflict graph from \cite{Jaf}). Along with the conflict hypergraph, the \textit{alignment graph} as defined in \cite{Jaf} was also used in \cite{PrL} to characterise index coding problems for which rate $\frac{1}{3}$ is feasible. Both of these graphs have the same vertex set, which is the set of messages $\cal W$. We recall these definitions as they will be used in this work also.
%
In \cite{PrL}, we developed a new framework for studying the rate $\frac{1}{3}$ feasibility of groupcast index coding problems. We recall the basic definitions and results from \cite{PrL}. We build on these results in this paper. 
%%%
\begin{definition}[Conflict hypergraph]
The conflict hypergraph is an undirected hypergraph with vertex set $\cal W$ (the set of messages), and its hyperedges defined as follows. 
\begin{itemize}
\item For any receiver $j$ demanding any message $W_k$, $W_k$ and $Interf_k(j)$ are connected by a hyperedge (shown dotted in our figures), which is denoted by $\{W_k,Interf_k(j)\}$.
\end{itemize}
%Two messages $W_i$ and $W_j$ are said to be \textit{in conflict} if there is a hyperedge connecting the two (other messages can be a part of this hyperedge).
%
\end{definition}
It was shown in \cite{PrL} that the conflict hypergraph sufficiently captures the essence of the index coding problem.
%%%%
\begin{definition}[Alignment graph and alignment sets - \cite{Jaf}]
The alignment graph is an undirected graph with vertex set $\cal W$ and edges defined as follows. The vertices $W_i$ and $W_j$ are connected by an edge (called an \textit{alignment edge}, shown in our figures by a solid edge) when the messages $W_i$ and $W_j$ are not available at a receiver demanding a message other than $W_i$ and $W_j$. A connected component of the alignment graph is called an \textit{alignment set}.
\end{definition}
%%%
\begin{definition}[Restricted Index Coding problem]
Let $\mathbb I$ denote an index coding problem with message set $\cal W$. For some ${\cal W}'\subseteq {\cal W}$, a ${\cal W}'$\textit{-restricted index coding problem} is defined as the index coding problem ${\mathbb I}_{{\cal W}'}$ consisting of 
(i) The messages ${\cal W}'$, (ii)The subset ${\cal T}_{{\cal W}'}$ of the receivers of $\mathbb I$ which demand messages in ${\cal W}'$, (iii) For each $j\in {\cal T}_{{\cal W}'}$ the demand sets $D_{{\cal W}'}(j)$ and the side-information sets $S_{{\cal W}'}(j)$ are restricted within ${\cal W}'$, i.e., $D_{{\cal W}'}(j)= D(j)\cap{\cal W}'$ and $S_{{\cal W}'}(j)= S(j)\cap{\cal W}'.$
%For the restricted index coding problem ${\mathbb I}_{{\cal W}'}$, it is easy to see that for any $j\in{\cal T}_{{\cal W}'}$ and for any message $W_k\in D_{{\cal W}'}(j)$, the \textit{restricted interfering set} $Interf_{k,{\cal W}'}(j)$ at the receiver $j$ for the message $W_k$ satisfies $Interf_{k,{\cal W}'}(j)=Interf_k(j)\cap{\cal W}'$.
\end{definition}
%%%

%%
The alignment graph and the alignment sets of the restricted index coding problem ${\mathbb I}_{{\cal W}'}$ are called the ${\cal W}'$-\textit{restricted alignment graph} and ${\cal W}'$-\textit{restricted alignment sets} respectively. A ${\cal W}'$-\textit{restricted internal conflict} is a conflict between any two messages within a restricted alignment set of ${\cal W}'$.

%%

%%%%
\begin{definition}[Triangular Interfering Sets, Adjacent Triangular Interfering Sets]
A subset ${\cal W}''\subset {\cal W}$ of size three is said to be a \textit{triangular interfering set} if all the messages in ${\cal W}''$ interfere simultaneously at some receiver, and at least two of the messages in ${\cal W}''$ are in conflict. Two distinct triangular interfering sets ${\cal W}_1$ and ${\cal W}_2$ are said to be \textit{adjacent} if ${\cal W}_1\cap{\cal W}_2=\{W_i,W_j\}$ such that $W_i$ and $W_j$ are in conflict. 

\end{definition}
%%%

%
%For example, in Fig. \ref{fig:rateonethird}, the sets of messages $\{W_1,W_3,W_4\}$ and $\{W_1,W_2,W_4\}$ are two adjacent triangular interfering sets, interfering at the receivers demanding $W_5$ and $W_6$ respectively.
%%%
\begin{definition}[Connected triangular interfering sets, Type-2 alignment sets]
\label{type2align}
Two triangular interfering sets ${\cal W}_1$ and ${\cal W}_2$ are said to be \textit{connected} if there exists a \textit{path} (i.e., a sequence) of adjacent triangular interfering sets starting from ${\cal W}_1$ and ending at ${\cal W}_2$. A \textit{type-2 alignment set} is a maximal set of triangular interfering sets which are connected to each other.
\end{definition}
%%%

The following theorem gives a necessary and sufficient condition for assigning vectors from a two dimensional space to the type-2 alignment sets. %(i.e., for satisfying the necessary condition given by Proposition \ref{type2sets2dim}).
%%%
\begin{theorem}
\label{norestrconflictstype2}
Let ${\cal W}'$ be a type-2 alignment set of the given index coding problem $\mathbb I$. If $\mathbb I$ is rate $\frac{1}{3}$ feasible, then ${\mathbb I}_{{\cal W}'}$ must be rate $\frac{1}{2}$ feasible which holds if and only if there are no ${\cal W}'$-restricted internal conflicts.
\end{theorem}
%%
%%\begin{IEEEproof}
%%The proof follows by combining the claims of Proposition \ref{type2sets2dim}, Corollary \ref{3cross1restrictedcorr} and Theorem \ref{restrictedratehalfthm}.
%%\end{IEEEproof}
%%%
%%Theorem \ref{norestrconflictstype2} is stricter than Theorem \ref{thmMAIS}, as the acyclic subset of messages of size $4$ in Theorem \ref{thmMAIS} is equivalent to a triangular interfering set within some type-2 alignment set with a restricted internal conflict. Theorem \ref{norestrconflictstype2} on the other hand requires  a complete type-2 alignment set to have no restricted internal conflicts, and is therefore more strict. We leave it to the reader to verify that the problem in Fig. \ref{fig:rateonethird} illustrates this difference.
%%\vspace{-0.25cm}%
%%
%\subsection{A new class of index coding problems with rate $\frac{1}{3}$ feasibility}
%\label{rateonethirdfeasibleproblems}
%We now prove the main result of this paper, which connects all the previously proved results and widens the class of index coding problems for which rate $\frac{1}{3}$ is achievable. Because of the framework we have developed in the previous subsections, the proof of this theorem is simpler than Theorem \ref{thmnocyclesforks}, while also subsuming that result.
%%%%
\begin{theorem} \label{thm:main}
A rate $\frac{1}{2}$ infeasible index coding problem $\mathbb I$ is rate $\frac{1}{3}$ feasible if every alignment set of $\mathbb I$ satisfies\textit{ either} of the following conditions.
\begin{enumerate}
\item It does not have both forks and cycles (a \textit{fork} is a vertex connected by three or more edges). %, and it is not a triangular interfering set.
\item It is a type-2 alignment set with no restricted internal conflicts.
\end{enumerate}
\end{theorem}
%%%
\section{A general necessary condition for rate $\frac{1}{3}$ feasibility} \label{sec:nec_condition}
We begin with the formal definition of strictly rate $\frac{1}{L}$ feasible and infeasible subsets. 

\begin{definition}[Strictly rate $\frac{1}{L}$ feasible and infeasible subsets] \label{def:strict_rate}
Let ${\mathbb I}$ be an index coding problem with message set ${\cal W}$. A subset ${\cal W}'\subseteq {\cal W}$ is said to be \textit{strictly rate} $\frac{1}{L}$ \textit{feasible} (or simply, \textit{strictly rate} $\frac{1}{L}$) if for any rate $\frac{1}{3}$ code given by an encoding function $\mathbb E$ (if it exists), we have $dim(V_{\mathbb E}({\cal W}'))=L$. A subset ${\cal W}'$ is called strictly rate $\frac{1}{L}$ infeasible if $dim(V_{\mathbb E}({\cal W}'))\neq L$ for any rate $\frac{1}{3}$ encoding function $\mathbb E$.
\end{definition}
%%%
Examples of strictly rate $\frac{1}{2}$ subsets include triangular interfering sets and type-2 sets. In the forthcoming part of this paper, we also construct other classes of strictly $\frac{1}{2}$ subsets as well as examples of strictly rate $1$ and rate $\frac{1}{3}$ subsets. Some of such sets help us to further characterize index coding problems which are rate $\frac{1}{3}$ feasible based on the existence of some substructures, like the type-2 set conditions given by Theorem \ref{norestrconflictstype2}. 
%%%
\subsection{Two Interference Configurations and their strictly rate $\frac{1}{L}$ feasible (infeasible) subsets} 
\label{subsec:inter_config}
Here, we will present two interference configurations (i) Successive triangular interference configuration (STIC) and (ii) Square pyramid interference configuration (SPIC). We will identify strictly rate $\frac{1}{L}$ feasible (and infeasible) subsets in these two configurations. We will also present an example based on two STIC subsets which is rate $\frac{1}{3}$ infeasible. 
%%%%%%
\begin{figure}[ht]
\centering
  \subfigure[STIC subset]{\label{fig:STICseta}\includegraphics[width=1.6in]{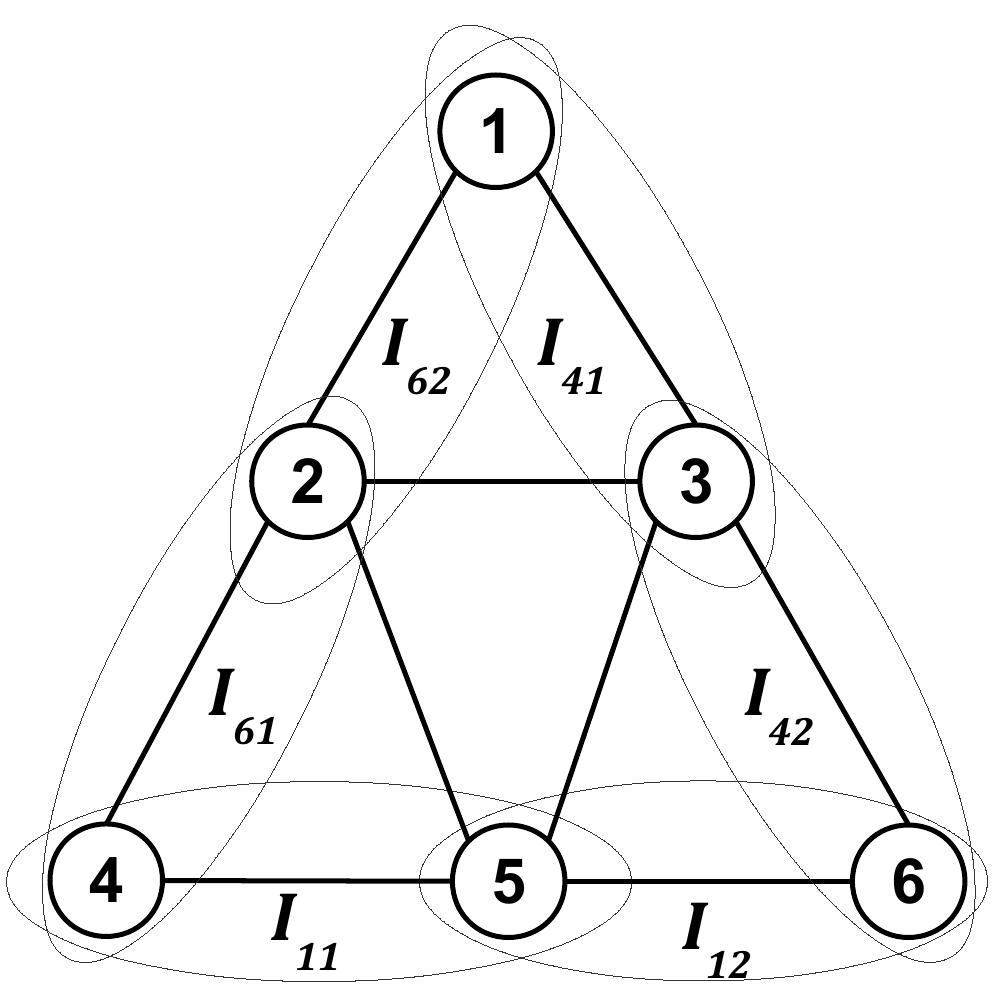}}
  \hspace{0.1in}
  \subfigure[Rate $\frac{1}{3}$ infeasible configuration]{\label{fig:STICsetDOUBLE}\includegraphics[width=1.7in]{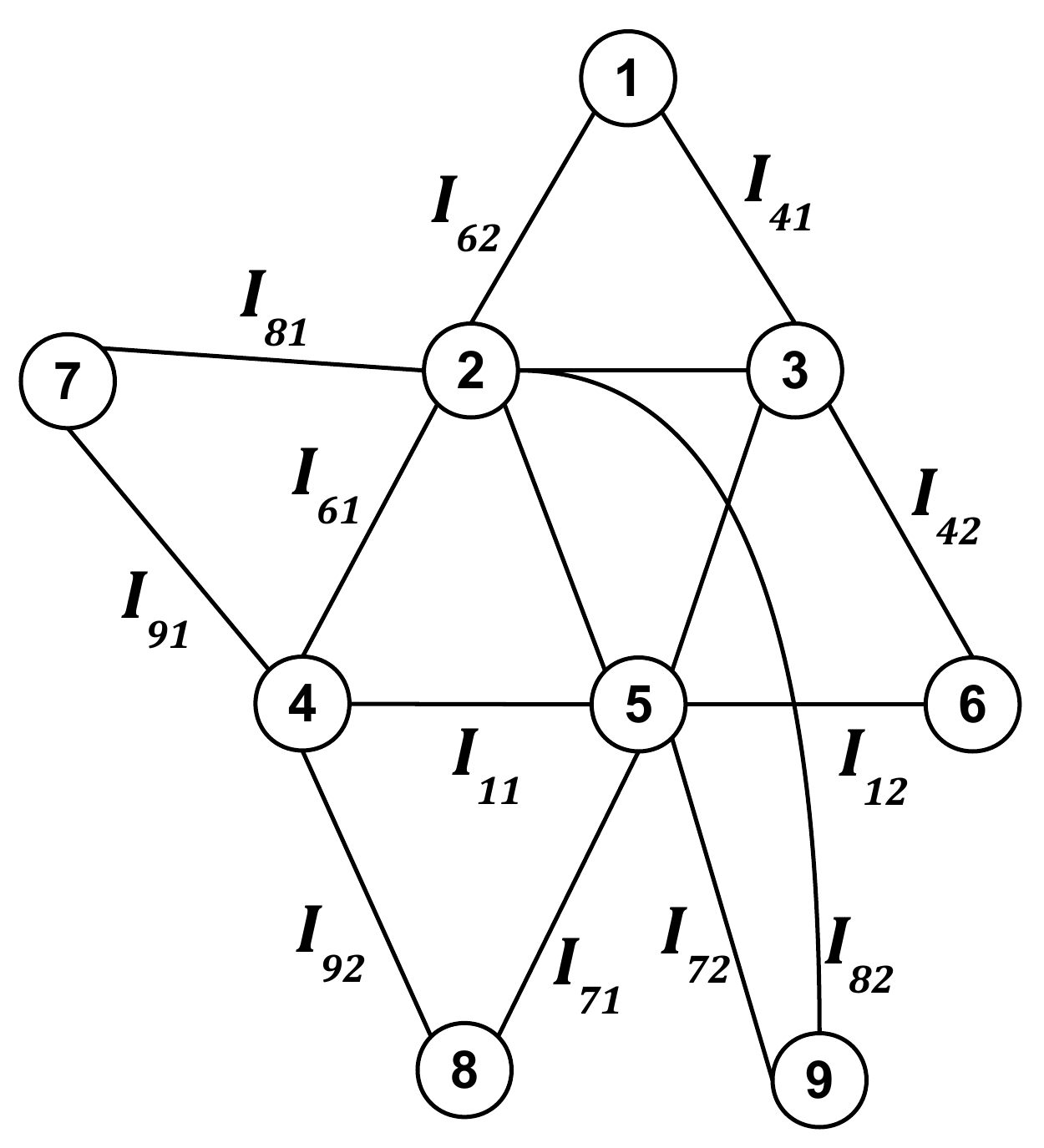}}
  \caption{Fig. \ref{fig:STICseta} shows a STIC subset. Messages in the ellipses $I_{ij}$ are interfering sets at some receiver demanding $W_i$. Fig. \ref{fig:STICsetDOUBLE} shows a rate $\frac{1}{3}$ infeasible configuration based on STIC sets. The sets $\{1,2,3,4,5,6\}$ and $\{7,4,2,8,5,9\}$ are both STIC sets. The messages along the edges marked $I_{ij}$ are interfering sets at a receiver demanding $W_i$.}
  \label{fig:STICset}
\end{figure}
%%%%%%%%
\begin{definition}[Successive Triangular Interference Configuration (STIC)]
A collection of messages ${\cal W}_{STIC}=\{1,2,3,4,5,6\}$ is said to be a \textit{successive triangular interference configuration} (STIC) subset (Fig. \ref{fig:STICseta}) if the following conditions hold.
\begin{itemize}
\item The sets $\{1,2,3\}$, $\{2,4,5\}$, and $\{3,5,6\}$ interfere at some receivers demanding messages other than $\{1,2,3,4,5,6\}$.
\item The sets $\{1,2\}$ and $\{2,4\}$ interfere at $W_6$, $\{1,3\}$ and $\{3,6\}$ interfere at $W_4$, and $\{4,5\}$ and $\{5,6\}$ interfere at $W_1$ (and there are no other conflicts amongst $\{1,2,3,4,5,6\}$).
\end{itemize}
\end{definition}

%\begin{figure}[ht]
%\centering
%\includegraphics[width=2in]{TRIANGLE_RATEONETHIRD_NONTYPE2.pdf}
%\caption{A STIC subset. Messages in the ellipses $I_{ij}$ are interfering sets at some receiver demanding $W_i$. The set of messages in each of the three corner triangles interfere at a receiver demanding some other message not in $\{1,2,3,4,5,6\}$}	
%\label{fig:STICset}
%\end{figure}
%%%
\begin{lemma}
\label{STIClemma}
Let $\{1,2,3,4,5,6\}$ be a STIC set as in Fig. \ref{fig:STICseta}. Then the following statements are true.
\begin{enumerate}
\item The set $\{2,3,5\}$ is strictly rate $\frac{1}{2}$ infeasible.
\item The sets $\{1,2,3\}$, $\{2,4,5\}$, and $\{3,5,6\}$ are strictly rate $\frac{1}{2}$ feasible. 
%\item Let ${\mathbb I}$ be an index coding problem which has for which  exists solutions to the STIC configuration  
\end{enumerate}
\end{lemma}
\begin{IEEEproof}
Proof of 1): Suppose $\{2,3,5\}$ is two dimensional for some rate $\frac{1}{3}$ index code given by $\mathbb E$. WLOG, let us assume that $V_2$ and $V_3$ are linearly independent, and thus $V_5\in sp(\{V_2,V_3\})$. Furthermore, $V_1\in sp(\{V_2,V_3\})$ as $\{1,2,3\}$ is interfering at some receiver. 

Now, if $V_5=\alpha V_2+\beta V_3$ for some $\alpha,\beta\in{\mathbb F}$ both non-zero, then it must be that $V_4=V_5$ (or a scalar multiple). Otherwise $V_4$ and $V_5$ will be independent, and hence we will have $sp(\{V_2,V_3\})=sp(\{V_2,V_5\})=sp(\{V_4,V_5\})$ and thus $V_1\in sp(\{V_4,V_5\})$, which is not allowed. Similarly, we should have $V_6=V_5$ too. This means that $V_4\in sp(\{V_3,V_6\})$, which is not allowed. Hence we cannot have $V_5=\alpha V_2+\beta V_3$ with $\alpha,\beta$ both non-zero.

Thus, WLOG, let $V_5=V_2$ (or equivalently, some constant multiple). Suppose $V_6$ (which is in $sp(\{V_3,V_5\})$ as $V_3$ and $V_5$ are independent) and $V_5$ are independent. Then $V_1\in sp(\{V_5,V_6\})$ which is not allowed. Thus $V_6=V_5=V_2$. However this means that $V_6\in sp(\{V_2,V_1\})$ for any choice of $V_1$, which is not allowed again. A similar contradiction arises if $V_5=V_3$. This concludes proof of 1).

Proof of 2): Clearly as the sets $\{1,2,3\}$, $\{2,4,5\}$, and $\{3,5,6\}$ are all interfering sets, we must have that each of them is strictly rate $\frac{1}{3}$ infeasible. Suppose $\{1,2,3\}$ is one dimensional. Then $\{2,3,5\}$ must also be one dimensional as $\{2,3,5\}$ cannot be two dimensional by 1). But then, $V_1=V_5$ which is not allowed. Thus $\{1,2,3\}$ must be two dimensional, and a similar argument applies for the other two sets.
\end{IEEEproof}
%%
%\begin{theorem}
%\label{strictlyfeasinfeas}
%Let ${\cal W}_1$ and ${\cal W}_2$ be subsets of the message set ${\cal W}$ of an index coding problem ${\mathbb I}$. Suppose there is a subset ${\cal W}_3\subseteq {\cal W}_1\cap {\cal W}_2$ such that the following is true.
%\begin{enumerate}
%\item In the ${\cal W}_1$-restricted index coding problem ${\mathbb I}_1$, the set ${\cal W}_3$ is strictly rate $\frac{1}{L}$ feasible (for some $1\leq L\leq 3$).
%\item In the ${\cal W}_2$-restricted index coding problem ${\mathbb I}_2$, the set ${\cal W}_3$ is strictly rate $\frac{1}{L}$ infeasible.
%\end{enumerate}
%Then the index coding problem ${\mathbb I}$ is rate $\frac{1}{3}$ infeasible.
%\end{theorem}
%%%%
%\begin{IEEEproof}
%Suppose ${\mathbb I}$ is rate $\frac{1}{3}$ feasible. Then, using the same allotment of vectors, one can construct rate $\frac{1}{3}$ index coding solutions for both ${\mathbb I}_1$ and ${\mathbb I}_2$. However, by the given conditions, this means that the set ${\cal W}_3$ is both $L$ dimensional and not $L$ dimensional, which is mutually contradictory. Hence ${\mathbb I}$ is rate $\frac{1}{3}$ infeasible.
%\end{IEEEproof}
%%%%
The following example, based on STIC sets, gives a rate $\frac{1}{3}$ infeasible problem based on an alignment structure constructed using two STIC sets, but does not have any triangular interfering sets (unlike all known previous examples). This also suggest the hardness of deciding the feasibility of rate $\frac{1}{3}$ index coding.%%%
\begin{example}
%%
%\begin{figure}[ht]
%\centering
%\includegraphics[width=2in]{TRIANGLE_RATEONETHIRD_NONTYPE2_EXTEND.pdf}
%\caption{The sets $\{1,2,3,4,5,6\}$ and $\{7,4,2,8,5,9\}$ are both STIC sets. The messages along the edges marked $I_{ij}$ are interfering sets at a receiver demanding $W_i$.}	
%\label{fig:STICsetDOUBLE}
%\end{figure}
%%%%%%
Consider an index coding problem ${\mathbb I}$ containing the alignment set structure shown in Fig. \ref{fig:STICsetDOUBLE}. There are two STIC sets  $\{1,2,3,4,5,6\}$ and $\{7,4,2,8,5,9\}$ which share the common set $\{2,4,5\}$. Consider the index coding problems ${\mathbb I}_1$ (respectively, ${\mathbb I}_2$) restricted to the STIC set $\{1,2,3,4,5,6\}$ (equivalently, $\{7,4,2,8,5,9\}$) and all the messages at which these messages interfere. Clearly, by Lemma \ref{STIClemma}, $\{2,4,5\}$ is strictly rate $\frac{1}{2}$ feasible in ${\mathbb I}_1$, while the same set is strictly rate $\frac{1}{2}$ infeasible in ${\mathbb I}_2$. Note that any rate $\frac{1}{3}$ feasible code for ${\mathbb I}$ is also feasible for ${\mathbb I}_1$ and ${\mathbb I}_2$. This means $\{2,4,5\}$ is simultaneously two dimensional and also not so. This is meaningless, hence we have that ${\mathbb I}$ is rate $\frac{1}{3}$ infeasible.  
\end{example}
%%%%%
\begin{remark}
As a general rule, if there is a subset of messages which is strictly rate $\frac{1}{L}$ infeasible in a restricted index coding problem of the given problem, and strictly rate $\frac{1}{L}$ feasible in another restricted problem, then the given problem must clearly be rate $\frac{1}{3}$ infeasible.
\end{remark}
%%%
We now define the SPIC sets, using which we obtain a new class of strict $\frac{1}{2}$ feasible sets.
%%%%%%%%%%%%%%%%%%%%
\begin{figure}[ht]
\centering
\includegraphics[width=1.5in]{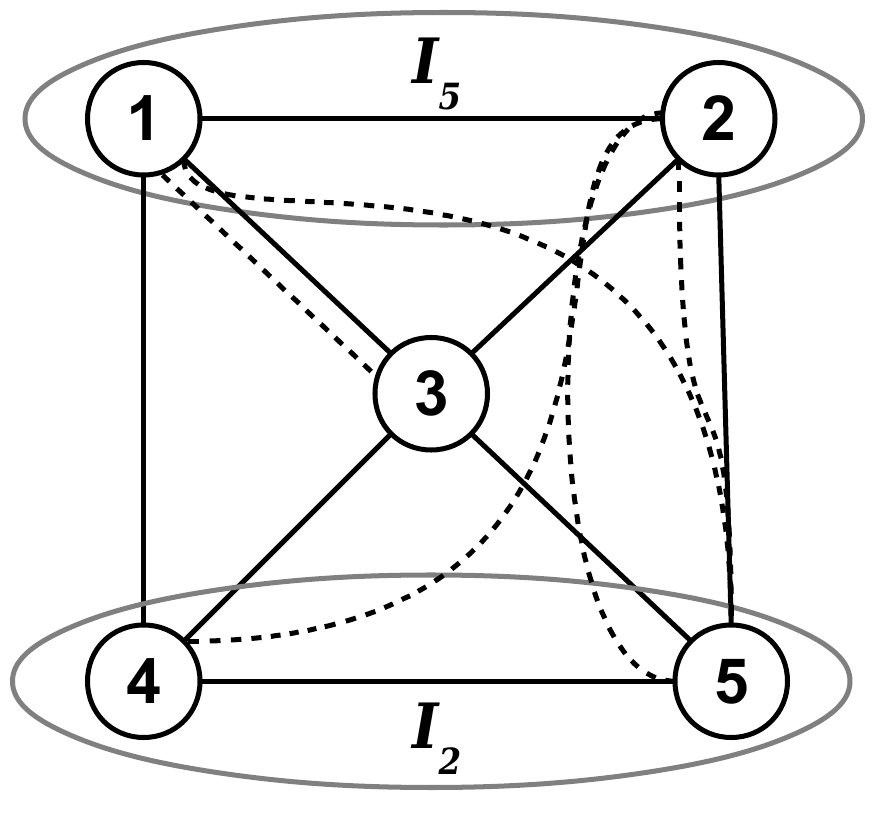}
\caption{A \textit{square pyramid interference configuration (SPIC)} set: The receivers, at which the sets $\{1,2,3\},\{1,3,4\},\{3,4,5\}$ and $\{2,3,5\}$ are conflicting, are suppressed. Definition \ref{squareinterferenceconfig} has the complete details.}	
\label{fig:squareset}	
\end{figure}
%%%%%
%%%
\begin{definition}[Square Pyramid Interference Configuration (SPIC)]
\label{squareinterferenceconfig}
A collection of messages ${\cal W}_{SPIC}=\{1,2,3,4,5\}$ is said to be a \textit{square pyramid interference configuration (SPIC)} set (see Fig. \ref{fig:squareset}) if the following conditions hold.
\begin{itemize}
\item Message sets $\{1,2,3\},\{1,3,4\},\{3,4,5\}$ and $\{2,3,5\}$ interfere at some receivers which do not demand the messages $\{1,2,3,4,5\}$.
\item Message set $\{1,2\}$ (denoted by $I_5$) interferes at a receiver demanding $W_5$, and $\{4,5\}$(denoted by $I_2$) interferes at a receiver demanding $W_2$. 
\item Messages $W_1$ and $W_3$ are in conflict. 
\item No conflicts other than the above exist between the messages $\{1,2,3,4,5\}$. 
\end{itemize}
\end{definition}
%%%
We now show that any SPIC set has to be strictly rate $\frac{1}{2}$, and also give other properties which it must satisfy, in the following lemma. %Further that the interference sets of size three containing the message sets $\{1,2\}$ and $\{4,5\}$ (in the ellipses of Fig. \ref{fig:squareset}) must also be strictly rate $\frac{1}{2}$. 
%%%
\begin{lemma}
\label{lemmaSPICconditions}
A SPIC set ${\cal W}_{SPIC}$ as shown in Fig. \ref{fig:squareset} satisfies the following conditions.
\begin{enumerate}
\item The triangular interfering sets $\{1,2,3\},\{1,3,4\},\{2,3,5\}$ and the set $\{1,2,3,4\}$ are strictly rate $\frac{1}{2}$.
\item ${\cal W}_{SPIC}$ is strictly rate $\frac{1}{2}$. 
\item The sets $\{1,4\},$ and $\{2,5\}$ are strictly rate $\frac{1}{2}$. 
\item The sets $\{4,5\}$ and $\{1,2\}$ must be strictly rate $1$.
\end{enumerate}
\end{lemma}
%%%
\begin{IEEEproof}
Proof of 1): This follows by Theorem \ref{norestrconflictstype2} as $\{1,2,3,4\}$ must be a subset of a type-2 alignment set, and also because $\{2,3,5\}$ is a triangular interference set. Thus, we must have $V_{\mathbb E}(\{1,2,3,4\})=V_{\mathbb E}(\{1,3\})$ for any rate $\frac{1}{3}$ feasible assignment $\mathbb E$. 

Proof of 2): Let $\mathbb E$ be the encoding function of some rate $\frac{1}{3}$ index code. Clearly $2\leq dim(V_{\mathbb E}({\cal W}_{SPIC}))\leq 3$. Suppose that $dim(V_{\mathbb E}({\cal W}_{SPIC}))=3$. By 1), we must therefore have that $V_5$ is independent of the space $V_{\mathbb E}(\{1,2,3,4\})$. Clearly, $V_2$ and $V_4$ must be linearly independent, as $W_2$ and $W_4$ are in conflict. By 1), we have $V_3\in sp(\{V_4,V_2\})$. It is easy to see that any such assignment for $V_3$ makes at least one of the sets $\{3,4,5\}$ and $\{2,3,5\}$ as three dimensional, which is not allowed. This proves 2).

%As we must have both $\{1,3,4\}$ and $\{3,4,5\}$ to be two-dimensional, we must therefore have $V_4=V_3$ (or a constant multiple). This means that $\{4,5\}$ is two dimensional, and hence we must have $V_2=\alpha V_1+\beta V_4 +\gamma V_5$, for some $\alpha,\beta,\gamma \in {\mathbb F}$ such that $\alpha\neq 0$. However, this means that $\{2,3,5\}$ (which is an interfering set at some receiver) is three dimensional, which is not allowed in any rate $\frac{1}{3}$ solution. This proves 2).

Proof of 3) and 4): By 2), it must be that $\{1,2,5\}$ must also be assigned vectors from the space spanned by $\{V_1,V_3\}$. Now, if $V_1$ and $V_2$ are linearly independent, then the conflicts will not be resolved at $W_5$. Thus, we must have $V_2=V_1$ (or a constant multiple). By the similar arguments, we must have $V_4=V_5=\alpha V_1+\beta V_3$ (for some $\alpha,\beta\in {\mathbb F}, \beta \neq 0$). This is the only assignment which is possible and it can be checked that 3) and 4) hold.
\end{IEEEproof}

\subsection{Necessary condition based on strictly rate $\frac{1}{2}$ feasible subsets} \label{subsec:nec_condition}

In the following theorem, we show that if the strictly rate $\frac{1}{2}$ subsets are appended together in a certain way, then the resultant subset must be strictly rate $\frac{1}{2}$. This can also be seen as a general necessary condition for rate $\frac{1}{3}$ feasibility based on the properties of strictly rate $\frac{1}{2}$ sets and their intersections.
%%%
\begin{theorem}
\label{thm:converse}
Let ${\cal W}_i:i=1,2,...,r$ ($r\geq 2$) be strictly rate $\frac{1}{2}$ sets of a rate $\frac{1}{3}$ feasible index coding problem ${\mathbb I}$ with message set ${\cal W}$, such that the following holds. 
%%%
\begin{itemize}
\item The sets $\left(\cup_{i=1}^{t-1}{\cal W}_i\right)\bigcap {\cal W}_t, 2\leq t\leq r,$ are strictly rate $1$ infeasible. 
\end{itemize}
%%%
Then the set $\cup_{i=1}^r {\cal W}_i$ must be strictly rate $\frac{1}{2}$. 

Thus, if the index coding problem ${\mathbb I}$ restricted to $(\cup_{i=1}^r{\cal W}_i)$ has any $(\cup_{i=1}^r {\cal W}_i)$-restricted internal conflicts, then ${\mathbb I}$ is not rate $\frac{1}{3}$ feasible. 
\end{theorem}
%%%%
\begin{IEEEproof}
If we are given that $\cup_{i=1}^r {\cal W}_i$ is strictly rate $\frac{1}{2}$, then the claim about rate $\frac{1}{3}$ infeasibility follows from the necessary condition for rate $\frac{1}{2}$ feasibility in \cite{Jaf}. Hence, it is enough to show that $\cup_{i=1}^r {\cal W}_i$ is strictly rate $\frac{1}{2}$. Let $\mathbb E$ be the encoding function of some $\frac{1}{3}$ index code. It is clear that $dim(V_{\mathbb E}(\cup_{i=1}^r {\cal W}_i))\geq 2$, as each set ${\cal W}_i$ is strictly rate $\frac{1}{2}$. We prove the claim by induction on $r$. The claim is true for $r=1$ by assumption that each set ${\cal W}_i$ is strictly rate $\frac{1}{2}$. Now suppose that the statement is true upto $r-1$. Then we have $\cup_{i=1}^{r-1} {\cal W}_i$ and ${\cal W}_r$ are strictly rate $\frac{1}{2}$ and $(\cup_{i=1}^{r-1} {\cal W}_i) \cap{\cal W}_r$ is strictly rate $1$ infeasible. This means that $(\cup_{i=1}^{r-1} {\cal W}_i) \cap {\cal W}_r$ must be strictly rate $\frac{1}{2}$ feasible as ${\cal W}_r$ is strictly rate $\frac{1}{2}$. Thus, we have $V_{\mathbb E}(\cup_{i=1}^{r-1} {\cal W}_i)=V_{\mathbb E}({\cal W}_r)=V_{\mathbb E}((\cup_{i=1}^{r-1} {\cal W}_i) \cap {\cal W}_r)$.
Thus $V_{\mathbb E}((\cup_{i=1}^{r-1} {\cal W}_i) \cup {\cal W}_r)=V_{\mathbb E}(\cup_{i=1}^{r-1} {\cal W}_i)$ as well, and hence $(\cup_{i=1}^{r-1} {\cal W}_i) \cup {\cal W}_r$ must be strictly rate $\frac{1}{2}$.
\end{IEEEproof}
%%%
%%%
Now, we will give three examples of sets (Corollary \ref{cor:type2},\ref{cor:xtype2}, and \ref{cor:spic}) which satisfy the condition in Theorem \ref{thm:converse}.

\begin{corollary}[Result from \cite{PrL}] \label{cor:type2}
Any type-2 alignment set of an index coding problem ${\mathbb I}$ is strictly rate $\frac{1}{2}$. 
\end{corollary}
\begin{IEEEproof}
Let ${\cal W}_i:i=1,...,r$ be the set of all triangular interfering sets in the given type-2 set. Clearly each one of them is strictly rate $\frac{1}{2}$. Furthermore, as the type-2 set is a maximally connected set of triangular interfering sets, we can assume WLOG that the sets ${\cal W}_i$ are in an order such that $(\cup_{i=1}^{t-1}{\cal W}_i)\cap {\cal W}_t, 2\leq t\leq r$ contains a conflict, and thus is strictly rate $1$ infeasible. The corollary follows by applying Theorem \ref{thm:converse}.
\end{IEEEproof}
%%%
%%%
\begin{definition}[Extended Type-2 Alignment Set]
Consider ${\cal W}_i:i=1,...,r$ be a maximal collection of type-2 alignment sets such that $(\cup_{i=1}^{t-1}{\cal W}_i)\cap {\cal W}_t, 2\leq t\leq r$ contains at least one conflict. Then the union $\cup_{i=1}^{r}{\cal W}_i$ is termed as extended type-2 alignment set (Xtype-2 set in short).
\end{definition}
%%%
We then have the following corollary, which we state without proof as it is a direct application of Theorem \ref{thm:converse}.
\begin{corollary} \label{cor:xtype2}
Any Xtype-2 set of an index coding problem ${\mathbb I}$ is strictly rate $\frac{1}{2}$. 
\end{corollary}

%%%
For our final corollary, we define SPIC alignment sets as follows, similar to type-2 alignment sets (Definition \ref{type2align}).
%%%%
\begin{definition}[Adjacent SPIC sets, SPIC Alignment Set]
Two SPIC subsets of ${\cal W}_1$ and ${\cal W}_2$ are said to be \textit{adjacent} SPIC subsets if ${\cal W}_1\cap {\cal W}_2$ is strictly rate $1$ infeasible. The SPIC sets ${\cal W}_1$ and ${\cal W}_2$ are called connected if there exists a sequence of adjacent SPIC sets from ${\cal W}_1$ to ${\cal W}_2$. A SPIC alignment set is a maximal set of SPIC subsets which are connected to each other. 
\end{definition}
%%%
%%%
\begin{figure}[ht]
\centering
\includegraphics[width=3.5in]{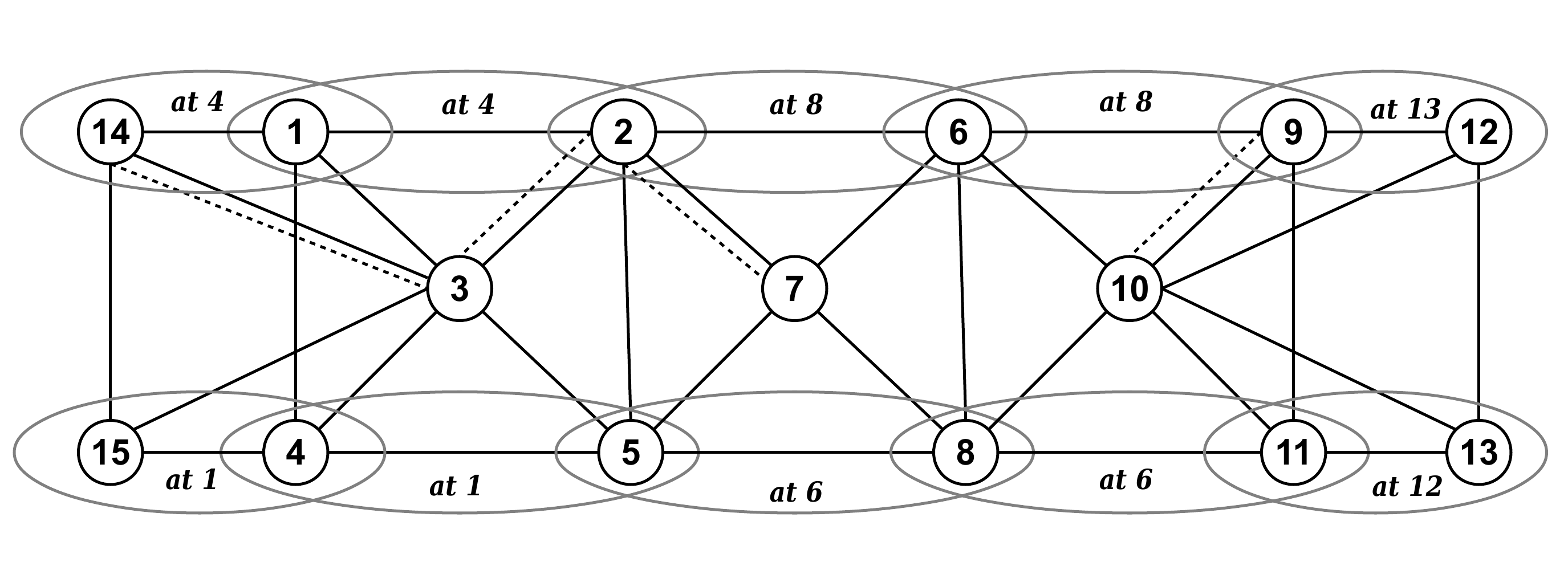}
\caption{A SPIC Alignment set. The sets $\{14,1,3,15,4\}$, $\{2,1,3,5,4\}$, $\{2,6,7,5,8\}$, $\{9,6,10,11,8\}$ and $\{9,12,10,11,13\}$ are all SPIC sets. The pair-wise intersections in the same sequence are $\{1,3,4\}$,$\{2,5\}$,$\{6,8\}$ and $\{9,10,11\}$ respectively, and all such intersections are strictly rate $1$ infeasible. The messages within the ellipses interfere at the messages as mentioned in the ellipse.}	
\label{fig:squaresetconnected}	
\end{figure}
%%%%%
Fig. \ref{fig:squaresetconnected} shows an example shows an example of a SPIC alignment set. We now give the related result.
%%%
\begin{corollary} \label{cor:spic}
A SPIC alignment set of ${\mathbb I}$ is strictly rate $\frac{1}{2}$. 
\end{corollary}
\begin{IEEEproof}
Let ${\cal W}_{a}$ be the SPIC alignment set. Let ${\cal W}_i:i=1,...,r$ be the set of all SPIC sets which comprise ${\cal W}_a$, ordered such that ${\cal W}_t$ is connected to $(\cup_{i=1}^{t-1}{\cal W}_i), 2\leq t\leq r$ (we can assume this WLOG as ${\cal W}_a$ is maximally connected set of SPIC sets). The corollary follows by applying Theorem \ref{thm:converse}.
\end{IEEEproof}

The sets described in Corollaries \ref{cor:type2}, \ref{cor:xtype2}, \ref{cor:spic} are only a few examples of the possible configurations of strictly rate $\frac{1}{2}$ sets in ${\mathbb I}$. We can imagine several other configurations as well, involving a mixture of Xtype-2 sets, SPIC sets, and other strictly rate $\frac{1}{2}$ sets (if they exist), satisfying the conditions in Theorem \ref{thm:converse}, leading to bigger structures which must be strictly rate $\frac{1}{2}$. An interesting open question is to  determine if  there a finite number of basic substructures, like the triangular interfering sets, SPIC sets, such that their maximally connected versions should satisfy specific properties for rate $\frac{1}{3}$ feasibility.
%%%%%%%%
\section{A Sufficient Condition for Rate $\frac{1}{3}$ Feasibility} \label{sec:suff_condition}
%%%
In this section, we introduce the notion of contraction of an alignment edge and contraction of an index coding problem. We prove that an index code for the contraction of an index coding problem can be extended to that of the original index coding problem. We also give a sufficient condition in terms of Xtype-2 sets in the maximal contraction of an index coding problem for the original index coding problem to be rate $\frac{1}{3}$ feasible.
\begin{definition}[Contraction of an alignment edge] \label{def:contraction}
An alignment edge connecting two vertices (messages) $W_i$ and $W_j$ which are not in conflict is said to be \textit{contracted} by identifying the vertices as a single vertex thereby making all the edges (alignment edges and conflict hyperedges) that were incident on $W_i$ and $W_j$ now incident on the newly created vertex. 
\end{definition}
%%%%
%%%%
When we contract any alignment edge, note that we get a derived index coding problem with one less the number of messages. The derived index coding problem is completely characterised by the contracted alignment graph and the correspondingly altered conflict graph. We can envision a sequence of such derived index coding problems, obtained by a sequence of contractions of the alignment set of the original problem, each with one lesser message than the previous. Abusing the definition, an index coding problem ${\mathbb I}'$ is said to be a \textit{contraction of} an index coding problem ${\mathbb I}$ if it is obtained by a finite sequence of contractions of the alignment edges of ${\mathbb I}$. The following definition captures the maximal cases after which we cannot contract anymore. 
%%%
\begin{definition}[Maximal Contraction of an Index Coding Problem]
Let $\mathbb I$ be a given index coding problem. Let ${\mathbb I}_{max}$ be an index coding problem obtained by a sequence of contractions of ${\mathbb I}$, such that any two messages of ${\mathbb I}_{max}$ connected by an alignment edge are also in conflict in ${\mathbb I}_{max}$. Then ${\mathbb I}_{max}$ is said to be a maximal contraction of ${\mathbb I}$.
\end{definition}
%%%
Clearly there could be multiple distinct maximal contractions of a given index coding problem, depending on the sequence of alignment edges which are contracted. 
%%%
\begin{lemma}
\label{lem:contract}
Let ${\mathbb I}'$ be a contraction of ${\mathbb I}$. Any scalar linear index coding solution of rate $R$ for ${\mathbb I}'$ can extended to a scalar linear index coding solution of rate $R$ for ${\mathbb I}$. 
\end{lemma}
%%%
\begin{IEEEproof}
WLOG, we assume that ${\mathbb I}'$ is obtained from ${\mathbb I}$ by a single contraction of an alignment edge. Let ${\boldsymbol W}=(W_1,...,W_n)$ denote the vector of messages in ${\mathbb I}$.

WLOG, consider that symbols $W_{n-1}$ and $W_n$ in ${\mathbb I}$ were contracted to get ${\mathbb I}'$. We denote the new symbol (vertex) which is created as the message $W'_{n-1}$. The vector of messages, the demand sets, the side-information sets and the interfering sets of ${\mathbb I}'$ (denoted as $\boldsymbol{W'}, D'(j), S'(j),$ and $Interf_j^{'}(k)$ respectively) can be obtained from those of ${\mathbb I}$ by replacing $W_i:1\leq i \leq n-1$ with $W_i':1\leq i \leq n-1$, and replacing $W_{n}$ with $W'_{n-1}$. We note that the number of receivers remains unchanged in ${\mathbb I}'$. 

Let ${\mathbb E}'$ be the encoding function of a solution to ${\mathbb I}'$. Thus we must have 
%%%
\begin{align}
\label{eqn100}
V'_k\notin V_{\mathbb E}'(Interf_j^{'}(k)),
\end{align}
%%%
for all $W'_k\in D(j),~\forall~\text{receivers}~j$, where $V'_k$ is the vector which is assigned to $W_k'$ in ${\mathbb E}'$. 

Consider the encoding function ${\mathbb E}$ for ${\mathbb I}$ defined as follows. 
%%%%%%%
\begin{align*}
V_k=
\begin{cases} V_k' & \forall 1\leq k\leq n-1.\\
V_{n-1}' & \text{if}~k=n.
\end{cases}
\end{align*}
%%%%%%%%%
We will now show that this assignment resolves all conflicts. Consider a message $W_k$ for some $k\leq n-2$ demanded at some receiver $j$. Thus $V_k=V_k'$. For some message $W_k$ demanded at receiver $j$, suppose $Interf_j(k)$ does not contain $W_{n}$. Then $Interf_j(k)=Interf_j^{'}(k)$ (except for the `dashes'). If $Interf_j(k)$ contains $W_{n},$ then $Interf_j^{'}(k)$ contains $W'_{n-1}$ in the place of $W_n$, apart from having other `dashed' messages corresponding to those in $Interf_j(k)$. In any case, we must have  
\begin{align}
\label{eqn101}
V_{\mathbb E}(Interf_j(k))=V_{{\mathbb E}'}(Interf_j^{'}(k)), \forall j,\forall k \leq n-2.
\end{align}
By (\ref{eqn100}) and (\ref{eqn101}), we have $V_k\notin V_{\mathbb E}(Interf_j(k))$. 

Now we consider $W_{n-1}$ (equivalently, $W_{n}$) demanded at receiver $j$. The receiver $j$ in ${\mathbb I}'$ demands the symbol $W'_{n-1}$, and it is given that (\ref{eqn100}) holds. Clearly, $Interf_j^{'}(n-1)$ does not contain $W'_{n-1}$. This means that $Interf_j(n-1)$ (equivalently, $Interf_j(n)$) does not contain $W_n$ or $W_{n-1}$. Thus, by similar arguments as above, we must have 
\[
V_{\mathbb E}(Interf_j(n-1))=V_{{\mathbb E}'}(Interf_j^{'}(n-1))
\]
(equivalently, $V_{\mathbb E}(Interf_j(n))=V_{{\mathbb E}'}(Interf_j^{'}(n-1))$). By assignment ${\mathbb E}$, we have $V_{n-1}\notin V_{\mathbb E}(Interf_j(n-1))$ (equivalently, $V_n\notin V_{\mathbb E}(Interf_j(n))$). Thus all the conflicts in ${\mathbb I}$ are resolved. This concludes the proof. 
\end{IEEEproof}
%%%

\begin{theorem} \label{thm:suff_condition}
A rate $\frac{1}{2}$ infeasible index coding problem $\mathbb I$ is rate $\frac{1}{3}$ feasible if there exists a maximal contraction $\mathbb I'$ of the index coding problem $\mathbb I$ such that the following conditions hold:
\begin{enumerate}[(a)]
\item Any Xtype-2  set in $\mathbb I'$ has no restricted internal conflicts. 
\item No three Xtype-2  sets have a message vertex in common.
\item For any distinct Xtype-2 sets ${\cal W}_i, {\cal W}_j$, if ${\cal W}_i \cap {\cal W}_j \neq \phi$, then there is no conflict between any two messages in ${\cal W}_i \cap {\cal W}_j$.
\end{enumerate}
\end{theorem}

\begin{IEEEproof}
To prove the theorem, we will give a solution to index coding problem $\mathbb I'$ and then apply Lemma \ref{lem:contract} to extend the solution to the actual index coding problem $\mathbb I$.
For notational convenience, we will drop the dash associated with the variables corresponding to the index coding problem $\mathbb I'$ (We will however refer to index coding problem as $\mathbb I'$). 
Let ${\cal W}_i, 1 \leq i \leq r$ denote the Xtype-2 sets of ${\mathbb I}'$. By assumption, we have ${\cal W}_i \cap {\cal W}_j \cap {\cal W}_k = \phi$, for all distinct $i,j,k$. Note that the condition (c) implies that there is no alignment edge between any two messages in ${\cal W}_i \cap {\cal W}_j$, as $\mathbb I'$ is a maximal contraction.

Consider a graph which has $r$ vertices, where each vertex represents an Xtype-2 set. There is an edge between two vertices in this graph, if the two Xtype-2 sets intersect in one or more messages. We will refer to this graph as the Extended Type-2 Intersection Graph (ETIG). We will first assign vectors to the edges in ETIG and then use that in turn to come up with an assignment for the messages in the index coding problem $\mathbb I'$. 

The algorithm to assign vectors to the edges of the ETIG is as follows. We start with the assumption that the edge set is non-empty, else the algorithm terminates straightaway. We repeat the below steps until all edges have been assigned vectors.
\begin{enumerate}
\item[1:] Pick an unassigned edge $e_{ij}$ between vertices $i$ and $j$. Let ${\cal V}_i$ denote the set of vectors already assigned to edges incident on vertex $i$ and similarly ${\cal V}_j$. 
\item[2:] Suppose both $sp({\cal V}_i)$ and $sp({\cal V}_j)$ are not $2$-dimensional, then assign a random $3\times 1$ vector to the edge.
\item[3:] Suppose exactly one of $sp({\cal V}_i)$ or $sp({\cal V}_j)$ is $2$-dimensional. For example, let $sp({\cal V}_i)$ be $2$-dimensional. Then, a random $3 \times 1$ vector from $sp({\cal V}_i)$ is assigned to the edge.
\item[4:] Suppose both $sp({\cal V}_i)$ and $sp({\cal V}_j)$ are $2$-dimensional. Then, a  $3 \times 1$ vector from the intersection $sp({\cal V}_i) \cap sp({\cal V}_j)$ is assigned to the edge. Such a vector always exists in the intersection since both $sp({\cal V}_i)$ and  $sp({\cal V}_j)$ are $2$-dimensional subspaces of a $3$-dimensional space.
\end{enumerate}

We note that at the end of the algorithm, each vertex with degree at least one in ETIG has a set of vectors associated with edges incident on them, the span of which is either $1$-dimensional or $2$-dimensional. An example ETIG and assignment of vectors to the edges in the graph is illustrated in Fig. \ref{fig:xtype2_vectors}.

\begin{figure}[ht]
\centering
  \subfigure[Xtype-2 Sets]{\label{fig:xtype2}\includegraphics[width=1.6in]{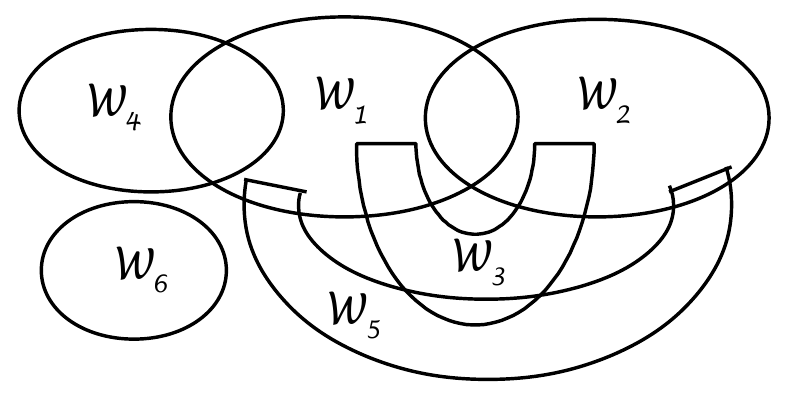}}
  \hspace{0.1in}
  \subfigure[ETIG of the Xtype-2 sets]{\label{fig:etig}\includegraphics[width=1.7in]{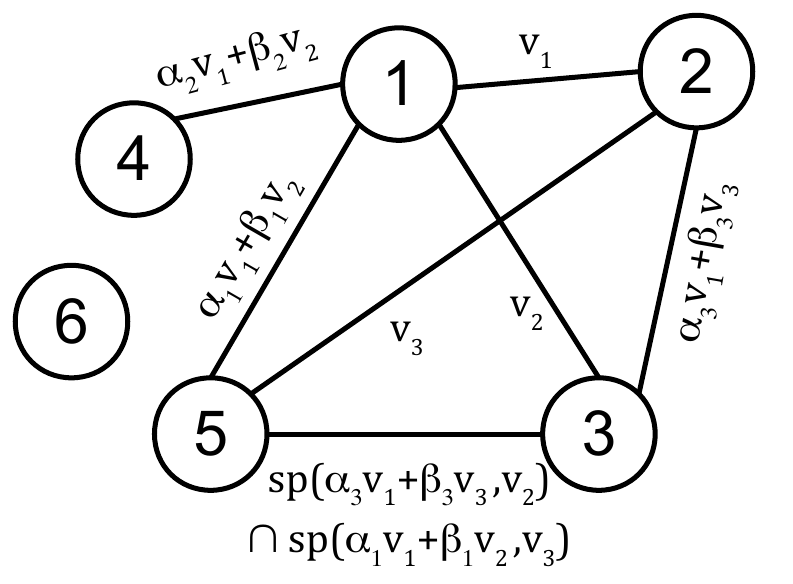}}
  \caption{The example has $6$ Xtype-2 sets with the intersections as shown. The ETIG of the Xtype-2 sets is given. To assign vectors, we start by assigning $v_1$ to edge $e_{12}$ and $v_2$ to edge $e_{13}$.  Then edges $e_{14}$ and $e_{15}$ are assigned random linear combinations of $v_1$ and $v_2$. Next, $e_{25}$ is assigned $v_3$. $e_{23}$ is assigned random linear combination of $v_1$ and $v_3$. Finally, since both vertices $3$ and $5$ are already assigned $2$-D spaces, a vector in the intersection of the two spaces is assigned to $e_{35}$. }
  \label{fig:xtype2_vectors}
\end{figure}

We now describe the procedure to assign vectors to the messages in the index coding problem $\mathbb I'$.
\begin{enumerate}
%\item If an alignment set is such that it has no three messages interfering at any receiver, then we assign an independently generated random $3\times 1$ vector (over a large field $\mathbb F$) to each message in the alignment set.
%\item If an alignment set is such that three messages interfere at any receiver, then there exists at least one Xtype-2 set contained within the alignment set. To these alignment sets, we assign vectors as follows:
%\begin{enumerate}[\hspace{-1cm} (a)]
\item Firstly, vectors are assigned to the messages in the intersections of Xtype-2 sets. We assign the same vector to all the messages in the intersection ${\cal W}_i \cap {\cal W}_j$. We pick the vector which was assigned to the edge which joined the vertices corresponding to $i^{\text{th}}$ and $j^{\text{th}}$ Xtype-2 set in ETIG and assign it to all the messages in the intersection ${\cal W}_i \cap {\cal W}_j$. This step is repeated for all intersections of Xtype-2 sets.
\item At the end of the above step, the messages in ${\cal W}_i$s maybe partially assigned. To assign vectors to remaining messages (if such non-assigned messages exist) in the Xtype-2 sets ${\cal W}_i$s, we consider the following three subcases:
\begin{enumerate}%[\hspace{-0.5cm} (i)]
\item For any $i$, if there is just one index $j$ such that ${\cal W}_i \cap {\cal W}_j \neq \phi$, then we have assigned only one vector to ${\cal W}_i$ at the end of step 1). Let the vector be denoted by $\underline{v}^{(i)}_1$. Now, we pick 
another $3 \times 1$ random vector $\underline{v}^{(i)}_2$. Then, we assign a random vector in the span of $\underline{v}^{(i)}_1$ and $\underline{v}^{(i)}_2$ to each message in ${\cal W}_i$.
\item For any $i$, if there is more than one index $j$ such that ${\cal W}_i \cap {\cal W}_j \neq \phi$, then we have assigned a $2$-dimensional space to messages in ${\cal W}_i$ at the end of step 1). To each of the remaining messages in ${\cal W}_i$, we assign a random vector from the same $2$-dimensional space.
\item For any $i$, if there is no index $j$ such that ${\cal W}_i \cap {\cal W}_j \neq \phi$ (corresponds to isolated vertices in ETIG), then we first pick a randomly generated $2$-dimensional space. Then, we assign a random $3\times 1$ vector in the $2$-D space to each message in the $i^{\text{th}}$ Xtype-2 set.
\end{enumerate}
\item For all other messages in ${\mathbb I}'$ (not in Xtype-2 sets), we assign a random $3\times 1$ vector.
\end{enumerate}
%%%
Let $\mathbb E$ denote the encoding function corresponding to this assignment and $V_k$ denote the vector assigned to the message $W_k$. We now show that this assignment resolves all conflicts in $\mathbb I'$, i.e., with high probability $V_k\notin V_{\mathbb E}(Interf_k(j))$. Consider a receiver $j$ which requests a message $W_k$. We will classify all the conflicts into the following cases and verify that conflicts are resolved in each case.
\begin{enumerate}[ (a)]
\item If  $|Interf_k(j)| = 1$, then the conflict is resolved, since any two messages which are in conflict are assigned linearly independent vectors according to the scheme. It is only the intersections of Xtype-2 sets which are assigned the same vector and we have already noted earlier that all the messages in any given intersection do not have any conflicts within them.
\item Consider the case when $|Interf_k(j)| = 2$. The two messages within the set $Interf_k(j)$ should have a conflict between them since $\mathbb I'$ is the maximal contraction. Now, we have the following two sub cases:
\begin{enumerate}[\hspace{-1cm} (i)]
\item Both $W_k$ and $Interf_k(j)$ belong to a Xtype-2 alignment set. This is not possible since then the Xtype-2 set will have a restricted internal conflict.
\item For any configuration of $W_k$ and $Interf_k(j)$ other than the case considered above, the conflict is resolved. This is true because any three messages which do not all belong to an Xtype-2 alignment set and which have conflicts between them are assigned three linearly independent  vectors according to the scheme. Here again, note that $Interf_k(j)$ cannot be contained within the intersection of two Xtype-2 sets (since all the messages in any given intersection do not have any conflicts within them).
\end{enumerate}
\item Consider the case when $|Interf_k(j)| \geq 3$. We note that in the maximal contraction $\mathbb I'$, any triangular interfering set of $\mathbb I'$ has to be contained within a type-2 alignment set and hence within a Xtype-2 set.
Thus, $Interf_k(j)$ is contained within a Xtype-2  set. Also since Xtype-2  sets of $\mathbb I'$ are assumed to have no restricted internal conflicts, $W_k$ has to be outside the Xtype-2 set. Since $Interf_k(j)$ lies within a Xtype-2 set and the messages within a Xtype-2  set are assigned vectors from a two dimensional space,  $V_{\mathbb E}(Interf_k(j))$ is a two dimensional space. Noting that $V_k$ is a random $3\times 1$ vector, we have that $V_k\notin V_{\mathbb E}(Interf_k(j))$ with high probability. Hence, the conflict is resolved.
\end{enumerate}
Thus, all the conflicts in $\mathbb I'$ are resolved. Now, the solution given above can be extended to give a solution of the original index coding problem $\mathbb I$ by 'uncontracting' one alignment edge at a time and doing an assignment as described in the proof of Lemma \ref{lem:contract}. This concludes the proof.
\end{IEEEproof}
%%%
\begin{remark}
We claim that Theorem \ref{thm:main} can be seen as a special application of Theorem \ref{thm:suff_condition} (which is the most general result known so far for rate $\frac{1}{3}$ feasible problems), and this can be shown using the proof details of Theorem \ref{thm:main} which is available in \cite{PrL}. We leave the details to the reader.  
\end{remark}
%%%

\end{document}